\documentclass[prb,twocolumn,aps]{revtex4}
\usepackage{epsf}

\begin{document}

\title{Comment to: ``Noncovalent functionalization of carbon nanotubes
by aromatic organic molecules'' [ Appl. Phys. Lett. 82, 3746 (2003) ] }
\author{P. Giannozzi}
\affiliation{NEST-INFM, Scuola Normale Superiore di Pisa,
Piazza dei Cavalieri 7, I-56126 Pisa, Italy}

\maketitle

Zhao {\em et al.} \cite{ddq} have calculated the interactions 
of various aromatic molecules with carbon nanotubes, using 
density-functional theory (DFT). They find evidence of 
noncovalent functionalization of carbon nanotubes, consisting
in a weak chemisorption with a small but sizable charge transfer 
from the nanotube to the molecule, or vice versa.
The charge transfer and associated change of the density of 
states at the Fermi level would explain the observed sensitivity 
of transport properties in carbon nanotubes to doping with
aromatic molecules. The same mechanism was previously invoked 
to explain the effects of adsorption of several simple gases 
(see Refs.[1-12] cited in Ref.[\onlinecite{ddq}], in particular 
Refs.[5,6,8,9]). 

The goal of this comment is to show that such mechanism is 
based on a shaky ground, and that Ref.[\onlinecite{ddq}] seems 
to ignore or to misinterpret recent work not supporting
the weak-chemisorption/charge-transfer picture.

Let us first focus on the case of O$_2$ adsorbed on 
the external wall of a nanotube.
Early work (Refs.[5,6,8,9] of Ref.[\onlinecite{ddq}]) 
used the Local-Density Approximation (LDA) -- whose
drawbacks, in particular the tendency to over-bind,
are well known -- and ignored spin polarization.
When spin polarization is properly taken into account,
however, very little charge transfer ($< 0.02e$) is
found\cite{noi,batraci}, even when using LDA. 
Calculations\cite{noi} based on the PBE flavor of the 
Generalized Gradient Approximation (GGA) yield very weak 
binding ($\sim 0.004$ eV) at large C-O distances 
($\sim 3.7$ \AA). With minor differences, the same pattern
holds for other flavors of GGA (Ref.[7] of Ref.[\onlinecite{ddq}]),
unless van der Waals (dispersion) forces are explicitely added\cite{batraci}.
Ref.[\onlinecite{ddq}] finds instead a sizable charge transfer
of $0.09 e$ at a much smaller C-O equilibrium distance of 2.8\AA\ 
with a binding energy of 0.1 eV, using GGA (no mention is 
done of the treatment of spin polarization). 

There is no direct experimental evidence, to the best of my knowledge, 
that the effects of O$_2$ adsorption on transport properties in carbon 
nanotubes are due to binding of O$_2$ to the wall of a perfect 
nanotube. Recent work points to a different mechanism (doping 
at the contacts) to explain the observed phenomenology\cite{avouris}.
More recent work shows that O$_2$ adsorption has no effect at all 
if impurities are carefully removed from nanotube samples\cite{goldoni}.

The results in Table I of Ref.[\onlinecite{ddq}] for both benzene 
and cyclohexane adsorbed on nanotubes are perfectly consistent with 
physisorption, as described by GGA\cite{vdW}: large equilibrium 
molecule-nanotube
distance, very small binding energy, no sizable charge transfer.
In spite of such evidence, Fig.4 of Ref.[\onlinecite{ddq}] is used 
to demonstrate hybridization of benzene $\pi$ states with the nanotube, 
as opposed to lack of hybridization in cyclohexane. Actually what is 
shown is that the top valence band includes states having $\pi$ 
character that are localized either on the nanotube or on benzene, 
but this does not imply that there is overlap or hybridization 
between the two sets of states: simply, their energies are in the 
same range.

\begin{figure}
\begin{center}
\epsfxsize=7.5cm
\vskip 0.5cm
\epsfbox{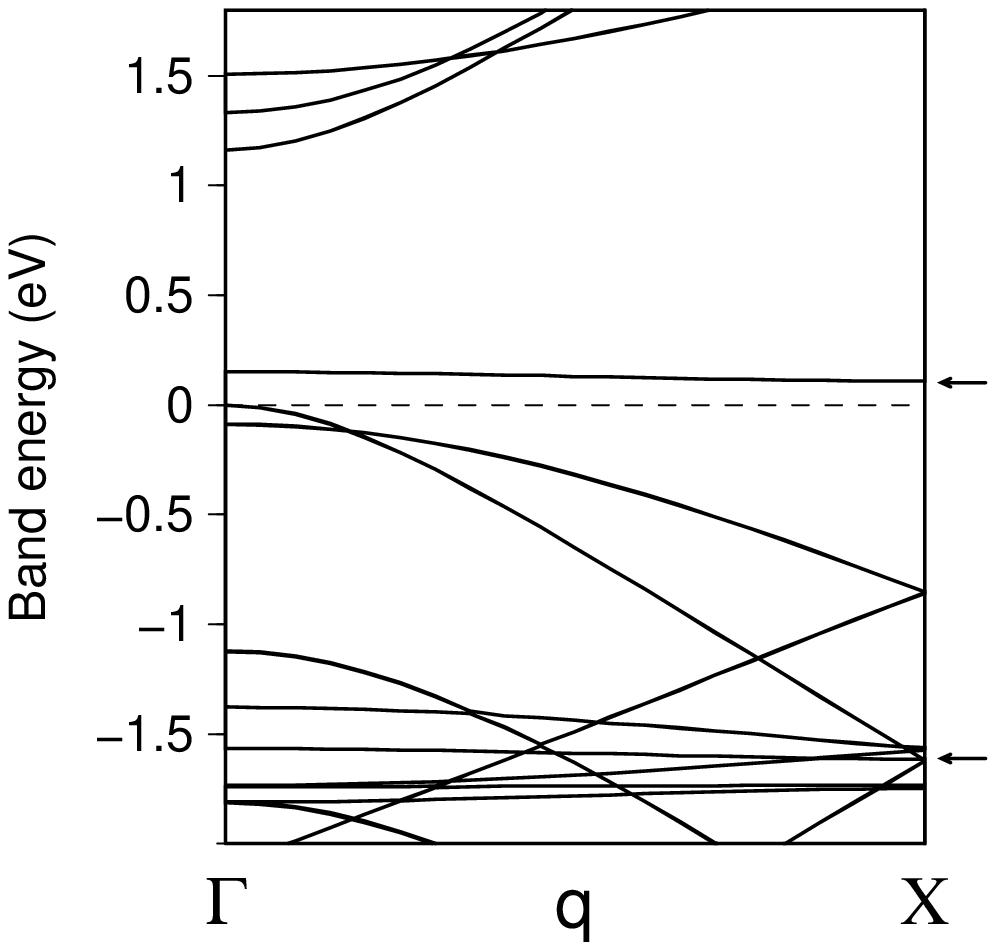}
\end{center}
\vskip -0.5cm
Fig.1. Band structure along the nanotube axis for a DDQ molecule at
5.3\AA\ from a (10,0) nanotube. The dashed line is the Fermi energy.
The supercell contains 80 nanotube atoms and a DDQ molecule, both
in their equilibrium structure. The arrows indicate the positions
of bands derived by the highest occupied and lowest unoccupied 
molecular state of DDQ, respectively.
\end{figure}

Fig.3 of Ref.[\onlinecite{ddq}] shows that for 
2,3-dichloro-5,6-dicyano-1,4-benzoquinone (DDQ) adsorbed on 
nanotubes, the Lowest Unoccupied Molecular Orbital (LUMO) state 
from DDQ at zone-boundary $X$ point lies just below the top of
the valence 
band of the nanotube. We remark however that even when
the DDQ molecule is placed at a large distance from the 
nanotube (5.3 \AA, a distance at which nanotube-molecule 
interactions are very small), the gap 
between the LUMO of DDQ and the top of the valence band 
of the nanotube is $\sim 0.1$ eV (see Fig. 1). 
The lineup between the top of the valence band of the nanotube 
and the LUMO of DDQ is thus mostly determined by the two systems 
separately and only marginally modified by their interactions. 
Whether such lineup can be reliably calculated with current
approximations to DFT is not obvious. If we assume that the 
occupied states of the two systems are correctly lined up,
the well-known underestimate of the gap between the Highest 
Occupied Molecular orbital (HOMO) and the LUMO will lead to
an incorrect lineup of unoccupied with occupied states.
For DDQ, the calculated HOMO-LUMO gap with GGA is 1.63 eV.

In conclusion, the results of DFT calculations in weakly bound 
systems like molecules adsorbed on nanotubes should be handled 
more carefully.

Partial support from MIUR grant PRIN 2001-028432 is acknowledged.

\end{document}